\documentclass[useAMS,usenatbib]{mn2e}
\usepackage{graphicx}
\usepackage{aas_macros}
\usepackage{amssymb}

\title[Multi-wavelength observations of the transitional millisecond pulsar 
binary  XSSJ12270-4859]
{Multi-wavelength observations of the transitional millisecond pulsar binary  
XSS\,J12270-4859
}
\author[D. de Martino et al.]{D.~de Martino$^{1}$\thanks{E-mail: demartino@oacn.inaf.it},
A.~Papitto$^{2}$, 
T.~Belloni$^{3}$, 
M.~Burgay$^{4}$, 
E.~De Ona Wilhelmi$^{2}$,
\newauthor
J.~Li$^{2}$,
A.~Pellizzoni$^{4}$,
A.~Possenti$^{4}$,
N.~Rea$^{2,5}$,
D.F.~Torres$^{2,6}$\\
$^{1}$ INAF $-$ Osservatorio Astronomico di Capodimonte, Salita Moiariello 16, I-80131 Napoli, Italy\\
$^{2}$ Institut de Ci\'ences de l'Espai (IEEC-CSIC), Campus UAB, c. Can Magrans s/n.,
E-08193 Barcelona, Spain\\
$^{3}$  INAF $-$  Osservatorio Astronomico di Brera, Via E. Bianchi 46, I-23807  
Merate (LC), Italy\\
$^{4}$ INAF $-$  Osservatorio Astronomico di Cagliari, Via della Scienza, I-09047
Serlagius (CA), Italy\\
$^{5}$ Anton Pannekoek Institute, University of Amsterdam, Science Park 904, 1098XH 
Amsterdam, The Netherlands\\
$^{6}$ Instituci\'o Catalana de Recerca i Estudis Avancais (ICREA), E-08010, Barcelona, Spain
}

\begin{document}

\date{Accepted 2015 September 8. Received 2015 September 8; in original 
form 2015 July 10}

\pagerange{\pageref{firstpage}--\pageref{lastpage}} \pubyear{2015}

\maketitle

\label{firstpage}

\begin{abstract}
We present an analysis of X-ray, Ultraviolet and optical/near-IR photometric data of the 
transitional millisecond pulsar binary XSS\,J12270-4859, obtained 
at different epochs 
after the transition to a rotation-powered radio pulsar state.  
 The observations, while confirming the large-amplitude
orbital modulation found in previous studies after the state change,
also reveal an energy dependence of the 
amplitudes as well as variations on time scale of months. 
The amplitude variations are anti-correlated in 
the X-ray and the UV/optical bands.
The average X-ray spectrum is described by a power law with 
$\Gamma$ index of 1.07(8) without requiring an additional thermal component. The
power law index $\Gamma$ varies from $\sim$1.2 to $\sim$1.0 
between superior and inferior conjunction of the neutron star.
We interpret the observed X-ray behaviour in terms of 
synchrotron radiation emitted in
an extended intrabinary shock, located between the pulsar and
the donor star, which is eclipsed due to the companion orbital motion. 
The G5 type donor dominates the UV/optical and near-IR emission 
and is similarly found to be  heated up  
to $\sim$ 6500\,K as in the disc state.
The analysis of optical light curves gives a  binary inclination 
 $46^o \lesssim i \lesssim 65^o$ and a mass ratio 
$0.11\lesssim q \lesssim 0.26 $.
The donor mass is found to be $0.15 \lesssim \rm M_{2} \lesssim 0.36\,M_{\odot}$ 
for a neutron star mass of 1.4\,M$_{\odot}$. 
The  variations in the amplitude of the orbital modulation are interpreted 
in terms of
small changes in the mass flow rate from the donor star.
The spectral energy distribution from radio to gamma-rays is composed by multiple 
contributions that are  different from those observed 
during the accretion-powered state.

\end{abstract}


\section{Introduction}

Millisecond pulsars (MSPs) are believed to be formed in binary systems containing 
old neutron stars (NSs), which are spun-up to very short periods during a previous
Gyr-long phase of mass accretion from an evolved companion. During this prolonged
phase the binary appears as Low-Mass X-ray Binary (LMXB) and when mass accretion
ceases a pulsar powered by the magnetic field rotation turns on and appears as
a radio and gamma-ray pulsar.  The MSP recycling
scenario \citep{Alpar82} found observational support through
the discovery of first MS radio pulsar \citep{Backer82} and has drammatically 
been confirmed by the detection of few hundreds of galactic radio MSPs in binaries. 
Furthermore a few tens of accreting NSs spinning at a few ms are 
known \citep{PatrunoWatts}, 
which are believed to be progenitors of MSP binaries. The transition between the
two states was first testified by PSR\,J1023+0038, a (1.67ms) MSP 
found to be in an accretion disc-state between 2000-2001 
\citep{Archibald09}. This object
was considered the "missing-link" between LMXBs and MSPs. 
It surprisingly turned on again in the X-ray band in 2013, showing X-ray
pulsations at the NS spin period and disc emission lines in the
optical band, thus testifying the presence of accretion at some level
\citep{Stappers13,Takata14,Patruno14,Bogdanov15}. 
Another striking case is the one of IGR\,J1825-2452, a binary in the M28 globular
cluster discovered  as an accreting MSP during an outburst in 2013 
\citep{Papitto13}, but identified as a radio pulsar few years before 
\citep{Begin06}. 
This source, turned into a radio pulsar at the end of the outburst, demonstrating
that transitions between the accretion-powered and rotation-powered states
can occur on timescales much shorter than secular evolution.
The hard X-ray source XSS\,J12270-4859 (henceforth XSS\,J1227)
was identified  as an unusual low-luminosity  
($\rm L_{0.3-10keV} \sim 6\times 10^{33}\,erg\,s^{-1}$ for a distance of 1.4\,kpc) 
LMXB, surprisingly associated to a high energy gamma-ray {\it Fermi}-LAT 
source \citep[][henceforth dM10,dM13]{deMartino10,deMartino13,Hill11}. The 
peculiar X-ray and gamma-ray emission from this source 
was explained in terms of a system hosting a fast rotating NS in a propeller
state \citep{Papitto14}. 
In late 2012/beginning 2013 it faded to the lowest X-ray \citep{Bogdanov14} 
and optical \citep[][henceforth dM14]{Bassa14,deMartino14} levels ever 
observed. Radio follow-ups revealed a fast (1.69\,ms) rotating radio 
pulsar \citep{Roy15}. Pulsed gamma-rays were also recently 
detected during the rotation-powered state \citep{Johnson15}. 
X-ray pulses at the NS spin could then be detected in the previous 
disc-accretion state \citep{Papitto15}, but not in the radio pulsar 
 state because it is too faint.   
These three examples open the challenging 
possibility that during the evolution  
there is a phase where the NS ''swings" back and forth from a radio pulsar
to a
sub-luminous ($\rm L_X \lesssim 10^{34}\,erg\,s^{-1}$) disc accretion state.
Analysing the spin period distribution of the accreting LMXBs,
the radio MSPs and those undergoing Type-I bursts, 
\cite{Papitto13} argued that nuclear
MSPs are in an earlier evolutionary stage, while accreting MSPs are in a phase
much closer to the switch-off and the turn-on of a radio MSP.
Thus, the transitions between the two states may reflect the interplay
between the NS magnetosphere and the mass transfer rate from the donor star
\cite[see][]{Stella94,Campana98,Burderi2001}.
It is also argued that some, possibly most, of the so--called ''redback" MSP 
binaries, which have non-degenerate donors, can display such state transitions
\citep{Roberts15}. Thus, all redbacks are potential ''swinging" MSPs and searches
for analogous sources are ongoing  \citep[e.g.][]{Bogdanov15a}.

\noindent The two known galactic 
field redbacks, PSR\,J1023+0038 and XSS\,J1227, linger in a sub-luminous
disc accreting state since mid-2013 and in a radio {pulsar} state since end-2012, 
respectively. They represent
ideal study cases to understand how transitions occur in redbacks.
We here present the analysis of observations in the X-rays and UV band 
of XSS\,J1227  during the current rotation-powered state 
acquired 
with {\it XMM-Newton} in 2014 June and in the optical with the {\it REM} 
telescope in 2015 January and February aiming at studying the 6.91\,h 
orbital variability of the source. We also compare the X-ray and 
UV/optical variability  observed in 2013 December 
when it was first found to have transited to a disc-free
state by \citet{Bogdanov14}, \citet{Bassa14} and dM14.
 We further analyse the spectral energy
distribution (SED) from radio to gamma-rays to identify the emission components
in the radio pulsar state.

\begin{table}
\flushleft
 \begin{minipage}{65mm}
  \caption{\label{obslog} Summary of the observations}
  \begin{tabular}{@{}lllll@{}}
  \hline
Telescope & Instrument &   UT Date  & UT Start & T$_{\rm expo}$ \\
          &            &            &          & (ks)   \\
\hline
{\it XMM-Newton} & EPIC-MOS1 & 2014-06-27 & 00:54:49 & 41.8 \\
                & EPIC-MOS2 & 2014-06-27 & 00:55:23 & 41.8 \\
                & OM-U      & 2014-06-27 & 01:03:50 & 42.5 \\
& & &  & \\
\hline
\end{tabular}

\begin{tabular}{@{}lllccc@{}}
  \hline
Telescope & Instrument &   UT Date  & UT Start & $\rm T_{expo}$ & \#  \\
          &            &            &          &  (sec)         &  \\
\hline
{\it REM} & ROSS2  & 2015-01-20 & 02:46 & 300 &  30 \\ 
    &       & 2015-01-21 & 02:40 & 300 &  31\\
    &       & 2015-01-23 & 02:32 & 300 &  37 \\
    &       & 2015-02-12 & 01:16 & 300 &  43 \\ 
    &       & 2015-02-13 & 01:24 & 300 &  44 \\  
    &       & 2015-02-14 & 01:10 & 300 &  51 \\  
    & REMIR & 2015-01-20 & 02:40 & 300$^{*}$ & 32\\
    &       & 2015-01-21 & 02:40 & 300 & 31 \\
    &       & 2015-01-23 & 02:24 & 300 & 38\\
    &       & 2015-02-12 & 01:20 & 300 & 44\\
    &       & 2015-02-13 & 01:10 & 300 & 51\\
    &       & 2015-02-14 & 01:44 & 300 & 46\\

& & & & & \\
\hline
\end{tabular}
$^{*}$ 60s exposures in dithering mode at 5 positions\\ 
\end{minipage}
\end{table}

\section[]{Observations}

\subsection{The XMM-Newton data}

XSS\,J1227 was observed twice with {\it XMM-Newton} after
the transition to the radio pulsar state which occurred between end 2012 and
beginning 2013 \citep{Bassa14}. The first observation, taken 
on Dec. 2013  December 29 (OBSID: 0727961401, hereafter Obs.~1), was analysed
by \cite{Bogdanov14}. The second pointing was performed   on 2014 June 27 
(OBSID:0729560801, hereafter Obs.~2) (see also Table\,\ref{obslog}). 
The former pointing used the EPIC-pn camera \citep{struder01} 
operated in Full Window Mode and with a thin
optical blocking filter. In such a configuration the instrument
retains a field of view of 25.6 x 26.2 arcmin, and has a temporal
resolution of 73.4 ms. During Obs.~2 the EPIC-pn was instead set to
the Timing fast mode to achieve a temporal resolution of 0.03\,ms with the
aim at detecting X-ray pulses at the 1.69\,ms rotational period 
of the NS. 
In this observing mode 2-D imaging capabilities are lost.  In
both observations a Full Window observing mode with a thin filter 
was used for the two
EPIC-MOS cameras  \citep{turner01}, giving a time resolution of 2.6 s.
Data were filtered to remove soft proton flares reducing the effective
exposure to 34.6 and 41.8 ks in Obs.~1 and 2, respectively. 
As discussed in \cite{Papitto15}, the source is not detected in 
the EPIC-pn data obtained in Timing mode during Obs.~2. Therefore
we focus only on the data acquired with the two EPIC MOS
cameras and draw comparison between the two observations.

\noindent We analyzed data using the XMM-Newton Science Analysis Software (SAS)
v.14.0\footnote{http://xmm.esac.esa.int/sas/}.
EPIC MOS source photons were extracted from a 40 arcsec wide circle
centered at the source position, to encircle $\approx 85\%$ of the
source photons. Background was accumulated from a 50 arcsec wide
circular region far from the source. X-ray photons were reported
to the Solar system barycentre using the optical position of the
source reported by \citep{masetti06}. XSS\,J1227 was detected by 
each MOS  camera at an average net level of 0.025(8)\,cts\,s$^{-1}$ and of 0.040(9)\,cts\,s$^{-1}$ 
in Obs.~1 and 2, respectively.

\noindent In Obs.~1 and 2, the Optical Monitor (OM) \citep{mason01}
was operated in Fast mode, giving a temporal resolution of 0.5\,s, 
with the U filter (3500-4800\,$\AA$). Ten exposures were acquired in both
observations  for a total of 32.7\,ks (Obs.~1) and 42.5\,ks (Obs.~2). The OM-U
band data of Obs.~1 were presented in \citet{Bassa14}. For a comparative analysis of
the two observations, the two data sets were reprocessed and 
photometry was extracted using the 
SAS task {\it omchain}. In a few exposures during Obs.~2 
the source was not detected and photometry was 
extracted performing manual detection of the 
target using the {\it omsource} task. 
The source was at an average magnitude of U=19.43(3) and 19.92(4) in 
Obs.~1 and 2, respectively. Correction to the Solar system barycentre was
also applied.

\subsection{The REM photometric data}

\noindent  XSS\,J1227 was observed from 2015 Jan. 20 to Jan. 23 (hereafter Run~1) and 
from 2015 Feb. 12 to Feb. 14 (hereafter Run~2) with the 0.6m INAF {\it REM}
 telescope 
in La Silla, Chile \citep{Zerbi04}.
The telescope is equipped with the ROSS2 camera\footnote{http://www.rem.inaf.it}
that performs
simultaneous exposures in the Sloan filters g', r', i' and z' and with
the REMIR camera \citep{Conconi04} covering simultaneously the near-IR in one
band. Integration times were 300\,s for all optical filters and a 
dithering of 5 exposures of 60\,s was used for the REMIR J-band exposures.
The night of Jan. 22 was not photometric and thus the data are not included here.
The total coverages were  3.2\,h, 3.51\,h and 4.1\,h in Run~1 and 
4.6\,h 5.3\,h 5.5\,h in Run~2.
The log of the photometric observations is also reported in Table\,\ref{obslog}.

\noindent  The photometric data sets were reduced using standard routines of 
IRAF to perform  bias and flat-field  corrections. Due to the low response
of the z' filter the corresponding images have not been analysed.
For the REMIR observations  the five dithered images were merged 
into a single frame. 

\noindent For each data set, aperture photometry was performed optimizing 
aperture radius and sky subtraction was done using annuli of different 
sizes.  Comparison stars  were used to check  and to correct for variable 
sky conditions.   The {\it REM}/ROSS2 and REMIR photometry was calibrated 
using the Sloan standard SA\,94\,242 observed each night whose near-IR
magnitudes are also tabulated in the 2MASS 
catalogue\footnote{http://www.ipac.caltech.edu/2mass/}.  
XSS\,J1227 was found at g'=18.62(2), r'=18.12(4), i'=17.96(2) 
in Run~1 and at g'=18.75(3), r'=18.28(3) and i'=17.91(4) in Run~2.
Because the source was barely detected in the individual J band images, image 
coaddition for each night was performed to provide a mean magnitude.  
XSS\,J1227 was at  J=16.98(4) and J=16.92(4) in Run~1 and 2, respectively.
The g', r' and i' light curves were also corrected to the Solar  system
barycentre.

\begin{figure}
\includegraphics[width=3.0in, clip=true, trim=0 5 0 5]{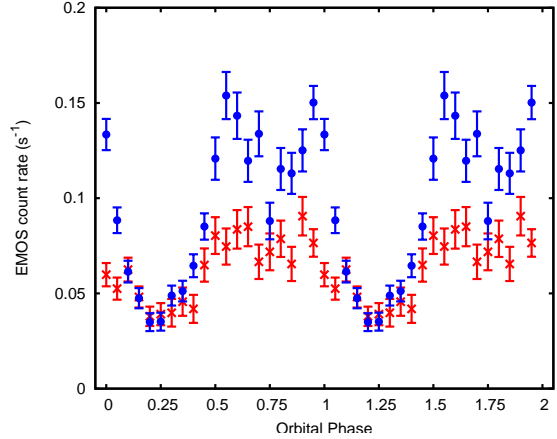}
\caption{X-ray orbital modulation observed during Obs.~1 (red
crosses) and Obs.~2 (blue circles), evaluated in 20 phase bins by
folding the background subtracted
 summed light curves observed by the two MOS cameras in the
0.3-10 keV band, at the 6.91\,h orbital period according to the radio
pulsar ephemeris given in \citet{Papitto15}. Phase 0 corresponds to the
passage of the NS at the ascending node of the orbit.}
\label{xrayorb}
\end{figure}

\begin{figure}
\includegraphics[width=3.0in, clip=true, trim=0 5 0 5]{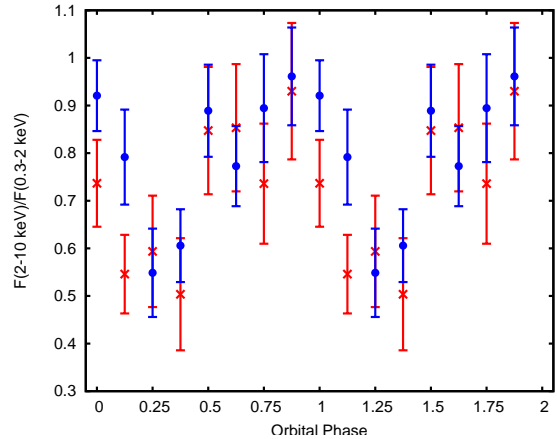}
\caption{Orbital modulation of the hardness ratio evaluated 
as the ratio of fluxes in the 2-10\,keV  and 
0.3-2 keV bands in 8 phase bins (see Fig.\,\ref{xrayorb}).}
\label{hr}
\end{figure}

\section{Results}

\subsection{The X-ray orbital variability}

The background subtracted summed MOS light curves in the 0.3-10\,keV range 
during Obs.~1 and 2 were folded 
at the orbital period using the radio pulsar ephemeris given
by \citet{Papitto15} where phase 0 is defined as the epoch at which the NS
passes at the ascending node of the orbit. The uncertainties on the determination
of the orbital phase with these ephemeris is 
$\lesssim 1\%$ for the two epochs.
The orbital modulation observed in Obs.~1 and 2 binned in 20 phase
intervals, each lasting $\simeq 1.25$ ks is
plotted in Fig.\,\ref{xrayorb}.

\noindent A minimum count-rate of 0.04\, cts\,s$^{-1}$ is
observed in both observations at orbital phase $\phi=0.25$, i.e. when
the donor star is between the pulsar and the observer. The intensity
then increases when the pulsar approaches the closest distance to the
observer ($\phi=0.75$). However, during Obs.~2 the increase of the
flux is steeper and the flux at maximum roughly doubles that during Obs.~1. 
The modulation amplitude, defined as $\rm (F_{max} - F_{min})/(F_{max} + F_{min})$, 
evaluated with a simple sinusoidal function is 0.25(4) and 0.69(6).
Also, a dip is observed during Obs.~2 centred at $\phi=0.75$ where the
count rate decreases by a factor $\sim$2.  This dip is instead
marginally visible in Obs.~1.

\noindent An orbital variability is also displayed by the 
hardness ratio evaluated as the ratio of the flux observed in the 2-10 keV band
to the flux in the 0.3-2 keV band (see Fig.\,\ref{hr}). 
The orbital modulation
was sampled using 8 phase bins in order to increase the counting
statistics of each point. In both observations, the hardness ratio increases by
a factor $\sim1.5$ at maximum of the modulation, implying  that 
the spectrum gets harder when the pulsar is at inferior conjunction.
The significance of the orbital variability of the hardness ratio is higher in
Obs.~2 than in Obs.~1. Modelling the orbital profile of the hardness ratio 
(in Fig.\,\ref{hr}) with a constant gives $\chi_{r}^2 = 1.76$ and 
$\chi_{r}^2 = 2.8$ for 7 d.o.f. in Obs.~1 and Obs.~2, respectively.  This gives
a probability that the profiles are compatible with a constant of 0.091 and 0.005,
respectively. The significance of a detection of the orbital variability in the 
hardness ratio is therefore 1.7 and 2.8$\sigma$ in Obs.~1 and Obs.~2. In both cases,
a sinusoidal fit to the orbital profile yields an 
improvement of the modelling with respect to a constant, which has a probability 
of $\sim 5\%$ of being due to statistical fluctuations (F-test). We also note that 
the source is significantly detected in all orbital phase bins in the two 
energy bands at a significance larger than $10\sigma$.

\subsection{The X-ray spectrum}

For each epoch, the spectra from the two MOS cameras 
in the range 0.3-10\,keV 
were fitted together using  the same
spectral parameters.  During Obs.~2 the spectrum is
well described ($\chi_r^2=1.18$ for 123 d.o.f.) by a power
law with index $\Gamma=1.07\pm0.08$ (all uncertainties on the spectral
parameters are quoted at the 90$\%$ confidence level).
Only an upper limit of
$N_H<5\times10^{20}$ cm$^{-2}$ (90$\%$ confidence level) could be set
on the absorption column. The 0.3-10 keV average unabsorbed flux is
$7.1\pm0.4\times10^{-13}$ erg cm$^{-2}$ s$^{-1}$. The addition of a 
thermal component, either a blackbody ({\sc bbodyrad} model in XSpec) 
or a more physical NS atmosphere ({\sc nsa} model in XSpec, \citep{Zavlin06}), 
does not give an improvement of the modelling. Fixing in {\sc nsa} 
the NS mass and radius to 1.4 M$_{\odot}$ and 10 km, and using the low field option
B=0\,G, we
obtain an upper limit on the 0.3-10\,keV 
flux of such a thermal component of
$2\times10^{-14}$ erg cm$^{-2}$ s$^{-1}$, i.e. $\simeq 3\%$ of the
total flux. 

\begin{figure}
\includegraphics[width=3.0in, clip=true, trim=0 5 0 5]{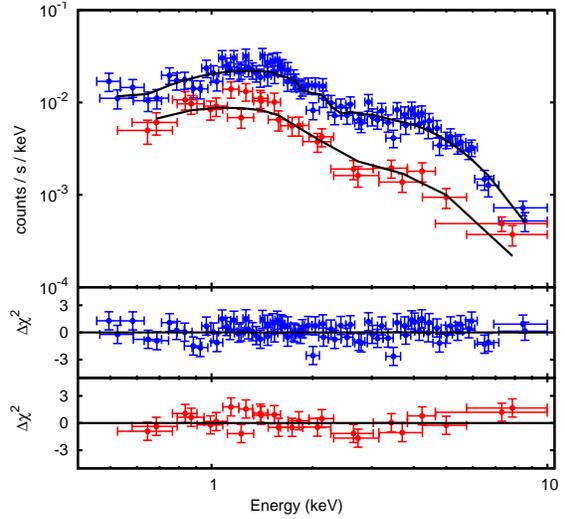}
\caption{The phase--resolved MOS spectra during Obs.~2 (top panel) 
at orbital phases close to the NS inferior conjunction ($\phi= 0.5-1.0$, blue
points) and close to the NS superior conjunction ($\phi= 0.1-0.4$, red points). 
Middle and bottom panels show residuals with the best fit absorbed power 
law model, shown as a black solid line.}
\label{xrayspec}
\end{figure}

\noindent We also modelled the spectra during Obs.~2 
at orbital phases close to the
NS superior conjunction ($\phi$ ranging between 0.1 and 0.4), and to
the NS inferior conjunction ($\phi$ ranging between 0.5 and
1.0). As indicated by the hardness ratio, close to the orbital 
minimum the spectrum gets softer and can be
described by a power law with index $\Gamma=1.30_{-0.15}^{+0.25}$ 
($\chi_r^2=1.03$ for 21 d.o.f.).
The 0.3-10 keV unabsorbed flux is $2.8\pm0.4\times10^{-13}$ erg cm$^{-2}$
s$^{-1}$. A thermal 
NS atmosphere component 
is not statistically required to model the spectrum, and
would contribute to at most to $40\%$ of the flux 
in the same energy range
($1.1^{+0.9}_{-0.4}\times10^{-13}$ erg cm$^{-2}$ s$^{-1}$). Close to
the NS inferior conjunction the spectrum is described 
by a harder ($\Gamma=1.0\pm0.1$) power law ($\chi^2_r=0.93$ for 73 d.o.f.) 
that emits a 0.3-10 keV flux of $1.08\pm0.08\times10^{-12}$ erg cm$^{-2}$
s$^{-1}$. In this case, a thermal component would contribute 
to at most 13\% of the
total flux. The fits to the MOS spectra and those at the two phases
using the simple absorbed power law are shown in Fig.\,\ref{xrayspec}.

\noindent The fit to the MOS spectra during Obs.~1 is
consistent with the results found
by \cite{Bogdanov14} for the EPIC-pn spectrum, the power law index being 
$\Gamma = 1.2\pm0.1$ ($\chi_r^{2}=0.87$ for
79 d.o.f.) and an upper limit to the hydrogen column density 
$\rm N_H<6.8\times10^{20}\,cm^{-2}$ (90$\%$ confidence level). The unabsorbed flux
in the 0.3-10\,keV range is $\rm 4.4\pm0.1 \times10^{-13}\,erg\,cm^{-2}\,s^{-1}$. 
Due to the lower sensitivity of the MOS cameras with respect to the EPIC-pn, 
we are unable to conclude on the detection of the NS thermal emission. Here
we note that the spectral fits to the EPIC-pn data do not improve but not
exclude a thermal component \citep{Bogdanov14}.

\noindent Comparing the power law indexes during the previous disc-accretion state 
(dM10,dM13) ($\Gamma$ = 1.7)
the spectra are harder during the disc-free states in Obs.~1 and 2 and the absorption
contribution is consistent with that previously determined in dM10 and dM13.
The luminosity in the 0.3-10\,keV range, assuming d=1.4\,kpc as derived 
from the radio  DM measures \citep{Roy15},  
results to be  1.0-1.7 $\rm \times10^{32}\,erg\,s^{-1}$
at the two epochs, one order of magnitude lower
than in the disc-accretion state.

\subsection{The UV/optical orbital modulation}

\begin{figure}
\includegraphics[width=3.0in, height=2.0in, clip=true, trim=0 5 0 5]{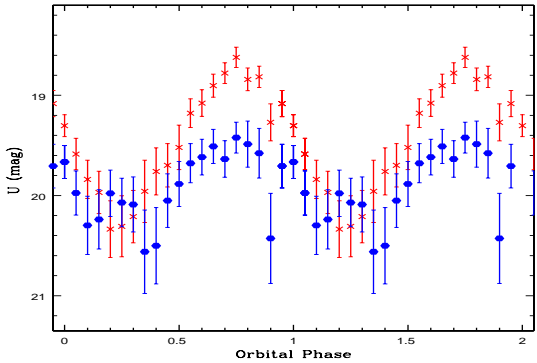}
\caption{Orbital modulation observed in the OM U band during Obs.~1 (red
crosses) and Obs.~2 (blue circles), evaluated in 20 phase bins.}
\label{uvorb}
\end{figure}

We folded the UV photometric data of Obs.~1 and 2 at the orbital period
using the same ephemeris adopted for the X-ray band. 
The orbital modulation in the UV band at the two epochs is reported in 
Fig.\,\ref{uvorb}. The UV flux is minimum at $\phi\sim0.25$ when the
source shows roughly the same magnitude (U$\sim$20.4) in both observations.
The UV flux reaches a maximum at $\phi \sim 0.75$ where it increases
by 1.43(4)\,mag and by 0.72(9)\,mag in Obs.~1 and 2, respectively. 
The slight decrease at $\phi \sim 0.9$ in Obs.~2 is not significant 
from inspection of the unfolded light curve.
The UV and X-ray band modulations are thus in phase, but the amplitudes 
in the two bands behave in opposite way. In Obs.~1 
the X-ray amplitudes are larger than in Obs.~2, but the UV amplitudes
are lower in Obs.~1 than in in Obs.~2 (see Fig.\,\ref{xrayorb}).

\begin{figure*}
\begin{tabular}{c}
\includegraphics[width=3.0in, angle=0, clip=true, trim=0 5 0 5]{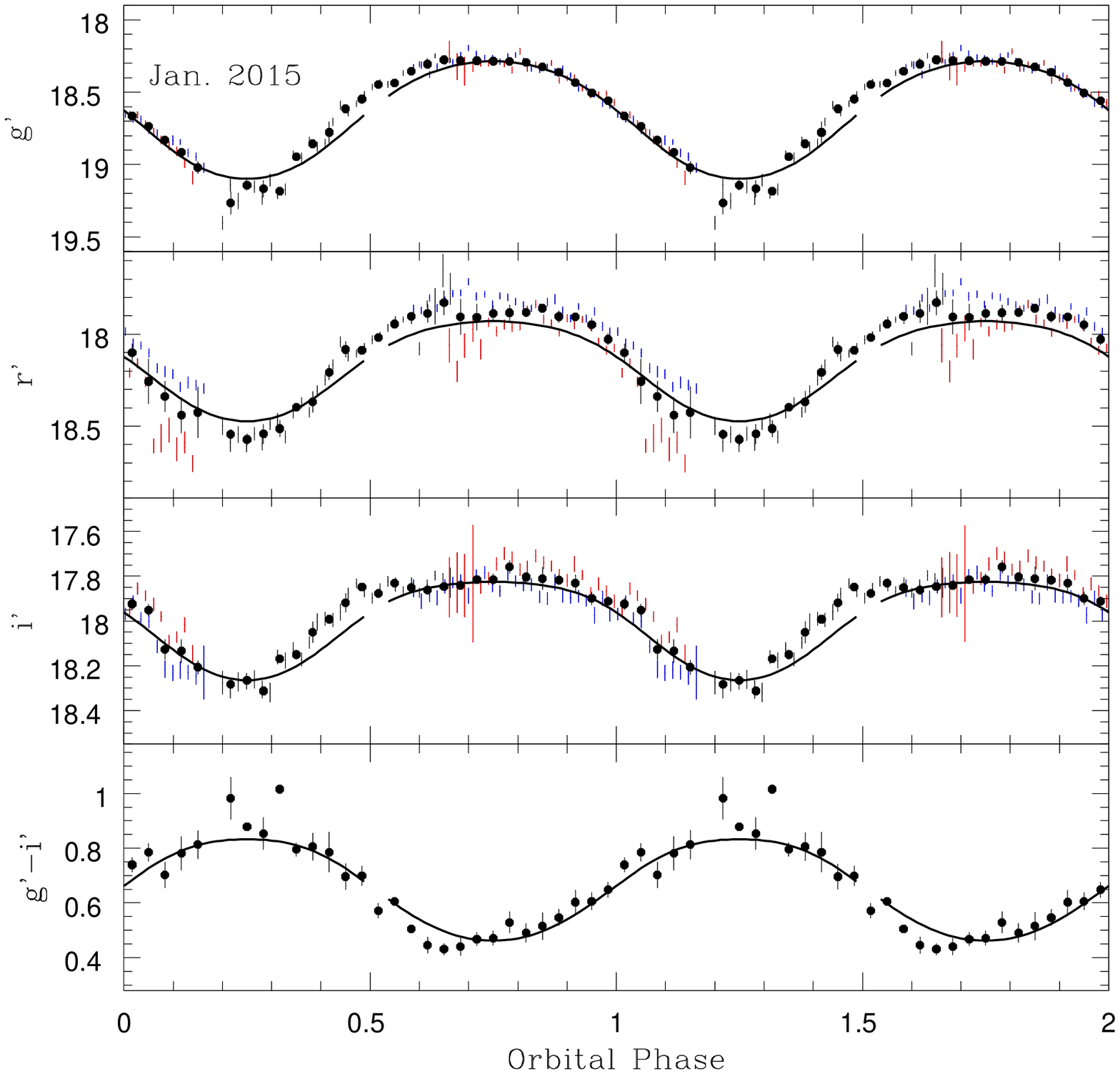}
\includegraphics[width=3.0in, angle=0, clip=true, trim=0 5 0 5]{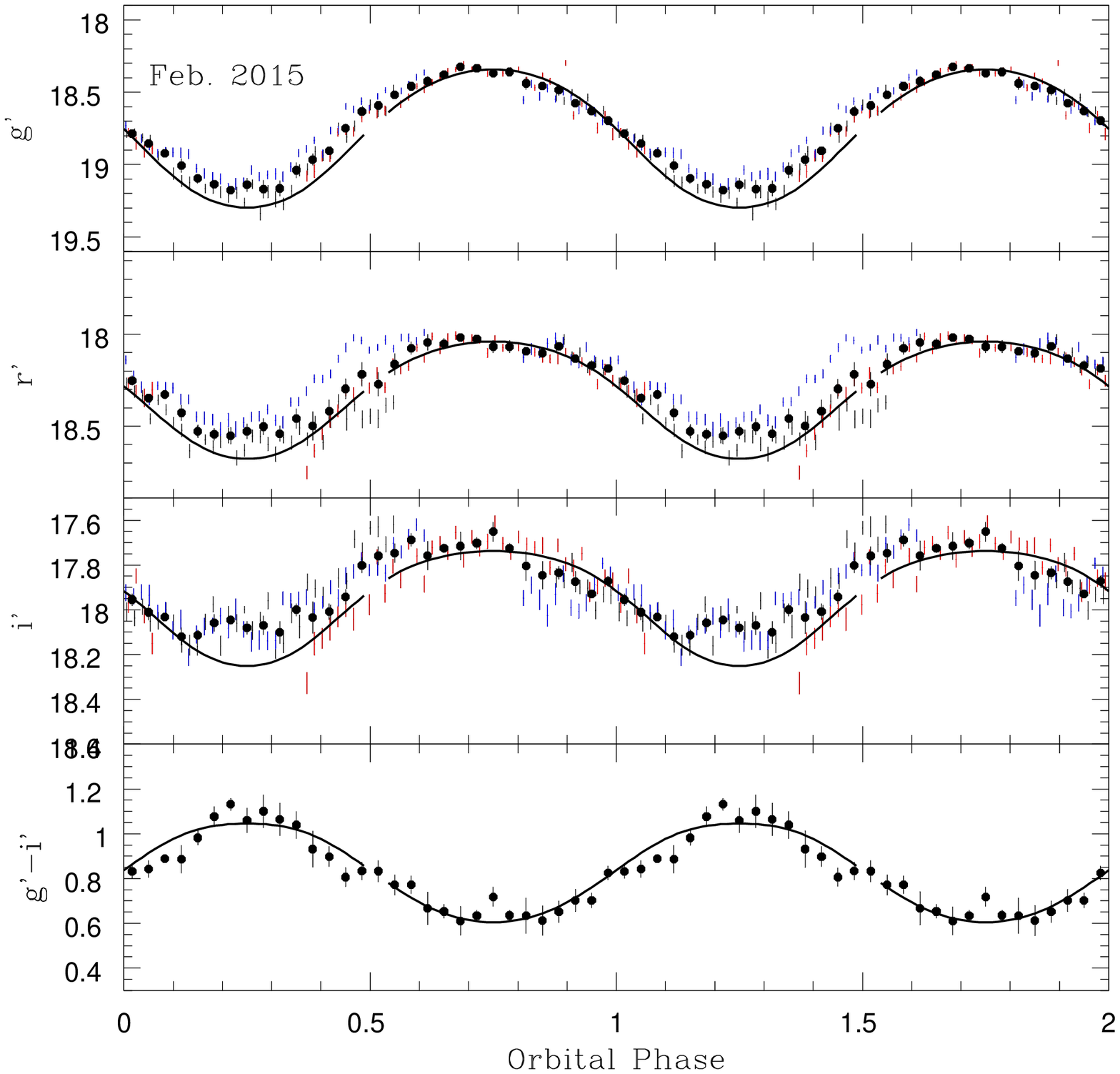}
\end{tabular}
\caption{From top to bottom: The orbital modulation in the optical 
g', r' and i' bands and the g'-i' index
as observed in Run~1 (left) and Run~2 (right). Single nights
 are reported with points of different colours and the average of the three 
nights as black points evaluated in 30 phase bins. The black lines report the
best fit {\sc Nightfall} curves  (see text for details). 
}
\label{optorb}
\end{figure*}

\noindent We folded the optical g', r' and i' {\it REM} 
photometry of 2015 Jan. (Run~1) 
and Feb. (Run~2) at the orbital period using the same ephemeris. 
In Fig.\,\ref{optorb} we report the
orbital modulation observed at the two epochs together with the g'-i' colour index.
Likewise the X-ray and U bands, the modulations in the optical/near-IR bands
show a minimum centred at $\phi$=0.25 and a maximum  at $\phi$=0.75. 
As evaluated with sinusoidal fits  the modulation amplitude 
is $\rm \Delta g'$=0.84(2)\,mag, 
$\rm \Delta r'$=0.74(1)\,mag and $\rm \Delta i'$=0.56(1)\,mag in Run~1, 
and it is $\rm \Delta g'$=0.72(2)\,mag, $\rm \Delta r'$=0.52(1)\,mag and 
$\rm \Delta i'$=0.40(1)\,mag in Run~2. The source is also on average 
brighter in g' and r'  by 0.1\,mag in  Run~1  than in Run~2.
A comparison with the optical level observed
in 2013 Dec. (dM14), the source was at about the same g' magnitude
as in Run~2, with a modulation of similar amplitude. 
The modulation is also colour dependent
and exemplified by the g'-i' index  (bottom panels of 
Fig.\,\ref{optorb}),  indicating a reddening at orbital minimum when the
donor star is at inferior conjunction.  The average values are g'-i'=0.66(6)
and g'-i'=0.84(4) in Run~1 and 2, respectively. The source thus appears
redder when on average fainter. 

\noindent The large amplitude modulation and strong colour variability confirm 
the previous finding that the donor star suffers irradiation (dM14).  
Similarly, we  applied the 
{\sc Nightfall}  code\footnote{The {\sc Nightfall} code is available
at http://www.hs.uni-hamburg.de/DE/Ins/Per/Wichmann/Nightfall.html} to fit
simultaneously the g', r' and i' light curves 
with the aid of the donor star radial velocity curve presented in dM14.
We adopted a blackbody temperature of 5000\,K as starting point for the 
unheated  star
and fixed the maximum allowed  temperature (500\,000\,K) 
for the irradiating star treated as a point source. The donor star is assumed
to fill its Roche lobe which is consistent with the occurrence of radio
eclipses for $\sim 40\%$ of the orbit \citep{Roy15}. We left the mass ratio  
$q$, the binary inclination $i$ and the donor star temperature $\rm T_{2}$ 
free to vary. We derive for Run~1 
$q=0.09\pm0.08$, $i =53^{+10}_{-13}$$^o$   and $\rm T_2$=6000$\pm$500\,K  
heated up to 6500\,K
($\chi_r^2$=9.5 for 362 d.o.f.) (uncertainties are at 1$\sigma$ confidence level)
and for Run~2  
$q=0.19^{+0.15}_{-0.19}$, $i = 50^{+20}_{-10}$$^o$ and 
$\rm T_2$=5900$\pm$400\,K heated up to 6300\,K 
($\chi_r^2$=15.3 for 482 d.o.f.). The
large $\chi_r^2$ are due to the small errorbars of the photometric measures.
The g', r' and i' light curves show asymmetries in the 
portion rising to the maximum but the inclusion of  cold or hot spots 
\citep{Romani12} does not improve the fits. Similar behaviour
is also found in the black widow binary PSR\,J1311-3438 by 
\cite{Romani15}, 
suggesting that the donor face is not fully heated by a point source and
that the asymmetric shock may provide at least some of the heating. 
We also used the U band photometry of Obs.~1 together with the donor star 
radial velocities and find $q =0.16^{+0.10}_{-0.05}$, $i =58^{+7}_{-12}$$^o$ 
and $\rm T_2$=5500$\pm$300\,K
heated up to 6000\,K ($\chi_r^2$=6.2 for 88 d.o.f.). 
The addition of the g' and r' photometry acquired in dec. 2013 (dM14) does
not improve the results because of the large scatter
and the sparse orbital coverage of the data. In all three data sets the 
mass ratio $q$ is consistent within errors 
with that derived from radio data by \citet{Roy15}, $q$=0.194(3).  
We note that the large filling factor of the
donor implies an aspherical shape for the star 
that causes strong temperature (brightness) variations over the surface
and the model code provides a mean temperature of the heated face 
of the secondary. 
The g', r', i' fluxes at the peak of the modulation observed in Run~1 and 2, when 
dereddened for intervening absorption $\rm E_{B-V}$ = 0.11
(dM10), are  consistent with a blackbody with temperatures 
similar to those derived from the light curve models, 6600$\pm$200\,K and 
6300$\pm$300\,K, respectively. 
The blackbody normalisation decreases by a factor of 1.17$\pm$0.01 
between Run~1 and 2.
We therefore confirm the findings in dM14, 
that the donor star temperature is compatible
with that of a G5 star seen at inferior conjunction and it reflects a F5 spectral
type at superior conjunction.
Since the binary parameters $q$ and $i$ cannot explain
the variability in the amplitudes observed at the different epochs, we
also conclude that the variations are due to changes in the visible area of 
the irradiated face of the donor star (see also Sect.\,4).
In the following we will assume $0.11\lesssim q \lesssim 0.26 $ 
and $46^o\lesssim i \lesssim 65^o$. For a NS mass of 1.4\,M$_{\odot}$, the
mass ratio gives a mass of donor star $0.15 \lesssim \rm M_{2} \lesssim 0.36$.


\section{Discussion}

We here discuss 
the X-ray and UV/optical orbital variability of XSS\,J1227 during
the current disc-free state. 

\subsection{The X-ray and optical variabilities}

\noindent The X-ray flux is strongly modulated and it is maximum close to the
inferior  conjunction of the NS.  The minimum occurs at the same phases of
the radio eclipses \citep{Roy15}, which are produced by the orbital motion of the
secondary star.  X-ray orbital modulations are also observed in other redbacks 
\citep{Roberts15}.
The X-ray spectrum as well as those at minimum and maximum
of the modulation are non-thermal. 
A dominant non-thermal spectrum may suggest that the X-ray emission 
originates in an intrabinary shock produced by the interaction of the
outflow from the donor star and the wind from the pulsar \citep{Arons_Tavani93}
and/or in the pulsar magnetosphere \citep{Possenti02}. 
The long duration of the X-ray eclipse, $\sim$ 30$\%$  of the orbital cycle, 
suggests that the X-ray emission is extended and originates close to 
the donor star and 
presumably at the L1 point, supporting an origin in the intrabinary shock. 
Similar interpretation is given for other eclipsing redbacks \citep{Bogdanov11,Li14},
where the X-ray
modulation is caused by the eclipse of the intrabinary shock 
emission by the donor. 
The eclipse depth and duration depend on the ratio between the donor star 
radius and the intrabinary shock separation and on the binary inclination.
The shock location or apex is defined as
$\rm r_s/a \sim 1/(1+\eta^{1/2})$ \citep{Arons_Tavani93},
where $\rm a$ is the binary separation and 
$\eta \sim \rm  \dot M\,v_w\,c/L_{sd}$
is the momentum ratio between the stellar wind and the pulsar wind.
Unfortunately, this quantity cannot be
determined because the mass loss rate during the radio pulsar
 state is unknown and
thus also the shock apex.  For what
follows we assume $\rm r_s \sim R_{L,NS} \sim 1.1-1.2\times 10^{11}$\,cm adopting
the mass of the components in Sect.\,3.3. 
For an isotropic pulsar wind the geometric fraction 
of the wind stopped by the shock, assumed
to be of the same size of the Roche-lobe filling donor star, is
$\rm f_s \sim 0.5\,(1 - cos\,\Omega) \sim 0.02$,
where $\rm \Omega$ = $atan$ $\rm (R_2/R_{L,NS})$ and 
$\rm R_2 = R_{L2} = 0.462\,(q/1+q)^{1/3}\,a \sim 3.2-4.2\times 10^{10}\,cm$
for the range of $q$ found in Sect.\,3.3.  This fraction is similar to those 
inferred in other redbacks \citep{Roberts15}.
 

\noindent The shock is expected to emit synchrotron radiation.
To estimate the synchrotron luminosity from the shock we 
follow \citet{Arons_Tavani93} and \citet{Kennel_Coroniti84}. Similar
arguments were also used by \citet{Stappers03} for the black widow 
binary PSR\,B1957+20 and first invoked for the redback binary 
PSR\,J0024-7204W in the globular cluster 47 Tuc by \citet{Bogdanov05} and
later also for PSR\,J1023+0038 by \citet{Bogdanov11}. 
The X-ray luminosity emitted by the shock depends on both the post-shock magnetic
field strength and the ratio of magnetic and particle energy density 
$\sigma$. This fraction is not known but could be as low as $\sigma$=0.003 
in winds dominated by kinetic energy, like the Crab pulsar, or as high as
$\sigma >>1$ for strongly magnetised winds \citep{Stappers03}. 
Given the small separation between the pulsar and the companion, 
the intrabinary shock is expected to be located in a strong magnetic field
\citep{Arons_Tavani93,Stappers03}. 
The magnetic field upstream of the shock is 
$\rm B_1 = [(\sigma/(1+\sigma)\,(\dot E/c\,f_p\,r^2)]^{1/2}$, 
where $\rm f_p$ is the pulsar isotropic factor and $\rm \dot E$ is the 
spin-down energy of the pulsar. 
In the two extreme cases,  assuming a pulsar energy loss of 
$\rm \dot E \sim 9\times 10^{34}\,erg\,s^{-1}$ \citep{Roy15}, an isotropic wind 
$\rm f_p$=1 and the shock at
 $\rm r_{s} \sim  R_{L,NS} \sim 1.2\times 10^{11}$\,cm,  we obtain
$\rm B_1 \sim$ 0.8\,G and $\sim$ 16\,G, respectively. The
 post-shock magnetic field is
$\rm B_2 \sim 3\,B_1 \sim$ 2.5\,G for $\sigma <<0.01$ and  it is
$\rm B_2 \sim B_1 \sim$ 16\,G for high-$\sigma$ values
\citep{Kennel_Coroniti84}.  
For
simplicity we assume that the observed X-ray luminosity 
totally originates via synchrotron emission from the shock. This
requires that the electrons are relativistic with 
$\rm \gamma \sim 2.5\times10^{5}\,(\epsilon/B_2)^{1/2}$, 
where $\epsilon$ is the observed photon energy, $\sim$ 10\,keV. Thus $\rm \gamma 
 \sim 2 - 5\times 10^{5}$ in the case of  magnetic and kinetic
energy dominated winds, respectively. 
The expected synchrotron luminosity 
is
$\rm L_{sync} \sim f_{sync}\, f_{\gamma}\,f_{s}\,L_{sd}$, 
where $\rm L_{sd}$ is the spin-down luminosity,   
$\rm f_{\gamma}$ is the fraction of spin down energy flux
that accelerates electrons with Lorentz factor $\gamma$ at 10\,keV, 
and  $\rm f_{sync} = (1+t_{sync}/t_{flow})^{-1}$ is the radiative efficiency 
of the synchrotron emission.  Here $\rm t_{sync}$
is the radiative synchroton loss time and  $\rm t_{flow}$ is the time of 
residence of particles radiating in the shock. 
For the assumed shock size of the order of the donor star diameter  
$\rm t_{flow}\sim (3r/c) \sim$ 6-9\,s for 
$\sigma <<1$ and $\rm t_{flow}\sim (r/c) \sim$ 2-3\,s for high-$\sigma$ flows 
\citep{Kennel_Coroniti84}. 
The radiative loss time 
$\rm t_{sync} =5\times10^{8}/(\gamma\,B_2^2)$ is $\sim$ 10\,s and 
$\sim$160\,s for
$\sigma >>1$ and $\sigma <<1$, respectively. Thus $\rm f_{sync}\sim 0.04-0.05$
for kinetic dominated winds and $\rm f_{sync}\sim$ 0.17-0.23 for magnetic dominated
winds. 
Using the geometric factor $\rm f_s \sim 0.02$ derived above and
the observed 0.3-10\,keV X-ray luminosity at orbital maximum of 
$\rm 2.5\times10^{32}\,erg\,s^{-1}$ , we obtain 
$\rm f_{\gamma} \sim 0.60-0.81$ for the highly magnetic case
and unphysical values $\rm f_{\gamma} \sim$ 3 for the 
kinetic dominated wind case. 
These large values reflect the geometrical uncertainties 
of the pulsar wind $\rm f_p$ and the fraction stopped by the shock 
$\rm f_s$, indicating that the pulsar wind is not isotropic and hence the
$f_s\gtrsim 0.02$ should be regarded as a lower limit. 
The unphysical values for $\rm f_{\gamma}$ in the low-$\sigma$ limit 
also imply that the wind must be magnetically dominated. Similar
conclusions were drawn for PSR\,J1023+0038  \citep{Bogdanov11}.


\noindent The X-ray spectrum hardens at orbital maximum with 
$\Gamma \sim 1.0$, whilst at minimum $\Gamma \sim 1.3$. A hardening at
orbital maximum is a new observational aspect in redbacks, as it has
been only detected in PSR\,J2129-0429 \citep{Hui15} so far. It is instead 
observed in other more  massive gamma-ray binaries
such as LS\,5039 \citep{Li11}. The hardening cannot be due to photoelectric
absorption since it would occur at orbital minimum. 
It could suggest  the presence of an additional contribution to 
the X-ray emission.
In spin-powered X-ray pulsars, a fraction $\sim 10^{-4}$ of the spin down 
luminosity is converted in X-rays.  
Using the empirical relation by \citet{Possenti02}, for 
$\rm L_{sd} \sim 9\times 10^{34}\,erg\,s^{-1}$ an
X-ray luminosity of magnetospheric origin 
$\rm \sim 3.5\times 10^{31}\,erg\,s^{-1}$  is 
expected in the 2-10\,keV range. 
This is not much different from that observed at minimum in the
same energy range:
$\rm L_{X,min} \sim 4.9\times 10^{31}\,erg\,s^{-1}$. 
This emission should be
pulsed at the spin period of the NS. Unfortunately the source is not detected 
in the EPIC-pn timing data during Obs.\,2, but \citet{Papitto15} 
derive an upper limit on the 0.5-10\,keV pulsed 
luminosity of $\rm \sim 1.6\times 10^{31}\,erg\,s^{-1}$.
The limit on the NS thermal emission at minimum
is estimated as $\rm \sim 0.4\, L_{X,min} \sim 2.5\times 10^{31}\,erg\,s^{-1}$
in the 0.3-10\,keV range (sect.\,3.2).  
Furthermore, at X-ray orbital minimum  the shock could still 
contribute, because of its large size  and  
the energy distribution of the electron population may be  softer in
the outer shock regions. 
Therefore, it is not possible to conclude whether the observed
softening at orbital minimum is due to the extended shock, to the NS or both.

\noindent Furthermore while only a hint of a dip superposed on the broad
maximum could be detected in Obs.1, this feature is observed when
source is brighter in Obs.2 (see Fig.\,\ref{xrayorb}). 
A double-peaked maximum  is is also present in  
PSR\,J2129-0429 \citep{Hui15,Roberts15}
at similar orbital phases. In PSR\,J1023+0038 an enhancement of emission is
observed at eclipse egress \citep{Bogdanov11}. These variations are
interpreted  as due to Doppler boosting due to 
the finite velocity of the shock wind in the direction of the post-shock flow
\citep{Arons_Tavani93,Stappers03,Bogdanov11,Huang12}. If the shock wraps the 
donor star, the plasma accelerates 
as it flows around the companion and, around inferior conjunction of the donor star,
the electrons are accelerated primarly in the direction of the observer. 
At this phases an increase of flux  would be  expected, but due to obscuration 
by  the donor star, the enhancements would occur before and after the 
eclipse. For PSR\,J1023+0038 the absence of enhancement prior the eclipse 
ingress was interpreted as due to a weaker outflow toward the leading edge of 
the eclipse \citep{Bogdanov11}.  
For post-shock velocities in the range 
$c/3 -c/3^{1/2}$, \cite{Arons_Tavani93} predict brightness variations 
by a factor of $\sim$1.3-2.2, respectively. 
The flux variations at the dip of a factor $\sim 2$
observed in Obs.~2 is consistent with this prediction.

\noindent In Obs.~2 the X-ray peak flux at maximum is  $\sim$2 times larger than on Obs.~1. 
If the mass loss rate and thus the luminosity  increases, 
 then  the momentum ratio
$\eta$ increases and the shock apex moves inward the NS Roche lobe.
However, a change by a factor of $\sim$2 in the mass loss rate 
will marginally  affect the apex radius, but it may  produce an 
increase in size of the shock. This would imply an increase of the
 time of residence  of electrons $\rm t_{flow}$ in the shock and of 
the fraction of the pulsar wind intercepted  
by the shock ($\rm f_{s}$). It could be also possible that the relativistic plasma 
accelerating as it flows around the companion may reach larger velocities 
giving rise a larger Doppler boosting effect as observed in Obs.~2.

\noindent The orbital modulation in the UV/optical is strong, amounting to
$\sim$ 0.7-0.8\,mag in the optical and up to $\sim$1.4\,mag in the UV. 
The optical colours along the orbital cycle indicate a reddening at inferior
conjunction of the donor star. The light curves are
well describe a G5 type donor star filling its Roche lobe and heated up to
$\sim$ 6500\,K, although the modelling considers a point-like 
irradiating source at the NS and not at L1.  
Thus the donor star is heated by high energy radiation from 
either the shock and/or by the pulsar wind. We here recall that irradiation of the
secondary  was also found to occur  before the state transition (dM14). 
As also noted in dM14, the donor is not expected to substantially cool, because
the thermal relaxation  time scale is of the order of the Kelvin-Helmhotz time
 of  the convective envelope, $\rm \tau_{KH} \sim (M_2^2/R_2\,L_2)$.  
The observed changes in the amplitudes of the 
UV/optical orbital modulation would then be due to changes of the 
visible area of the heated face of the
companion rather than to temperature variations. 
Thus, the opposite behaviour seen in the X-ray and UV modulation amplitudes 
could be explained in terms of a larger size of the shock, producing a stronger
X-ray modulation, and at the same time preventing the irradiated face 
to be totally visible, yielding  to a weaker modulation in the UV and 
optical bands. 
The heating effect on the donor star, 
$\rm L_{irr} = 4\,\pi\,R_{2}^2\,\sigma\,T_{irr}^4
\sim 1.2 - 2.2\times 10^{33}\,erg\,s^{-1}$ for $\rm T_{irr} \sim$
6500\,K\footnote{A detailed and rigorous treatment of the irradiation of the
donor star can be found in \citet{Ritteretal00}.
The effect of irradiation is to suppress the temperature gradient in 
the outer stellar envelope
and blocking the outward transport of energy through the face. Thus the emergent
flux from the irradiated face is: 
$\rm F_{irr} = \sigma T_{irr}^4 = W\, F_{inc}$, with W$\sim$0.5, the albedo for a
star in convective equilibrium.}, is larger than 
that estimated for the shock $\rm \sim 3\times 10^{32}\,erg\,s^{-1}$ 
in the 0.3-10\,keV range. However, if the X-ray spectrum extends well 
beyond 10\,keV, as 
observed by {\it NuSTAR} in PSR\,J1023+0038 with no spectral break up to 79\,keV
\citep{Tendulkar14}, the X-ray luminosity at orbital maximum 
in the 0.3-100\,keV range can be about one order of magnitude higher than 
that measured
in the {\it XMM-Newton} EPIC range. In this case the shock could be 
totally responsible for the heating of the donor star. 
If instead we consider  the pulsar spin-down power, 
we infer that $\sim 0.45$ of the spin-down luminosity
goes into the heating of the donor star. 
Observations at higher energies will clarify this issue. 



\begin{figure*}
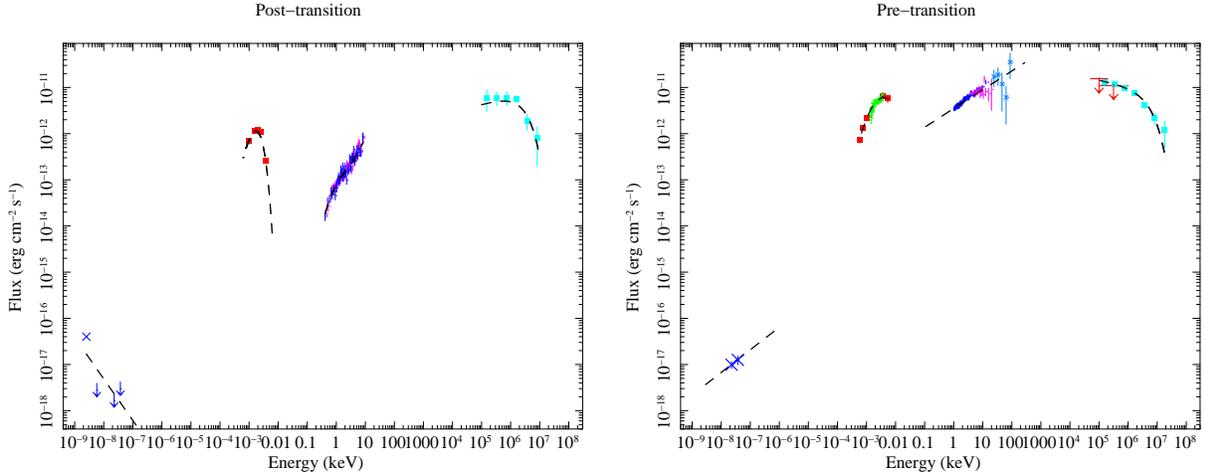

\includegraphics[width=2.5in, angle=-90, clip=true, trim=0 5 0 5]{fig6a.ps}
\includegraphics[width=2.5in, angle=-90, clip=true, trim=0 5 0 5]{fig6b.ps}
\caption{The broad-band spectral energy distributions (SED) of 
XSS\,J1227 during the rotation-powered
state (left panel) and during the sub-luminous pre-transition state (right panel).
The post-transition SED is constructed 
with the X-ray spectra of the two MOS cameras 
(blue and magenta), the  UV OM and optical REM photometry (red squares), the radio
measures reported by \citet{Roy15} (blue crosses) and the upper limit by
\citet{Bassa14} (blu arrows) and 
the {\it Fermi}-LAT (light blue)  fluxes reported in
\citet{Xing14} together with the corresponding fitted model (see text for
details). The pre-transition SED is taken from dM13 but substituting the {\it Fermi}-LAT
fluxes reported in the second release catalogue with those in \citet{Xing14}
together with the corresponding model fit.}
\label{seds}
\end{figure*}

\subsection{The spectral energy distribution}

We constructed a broad-band spectral energy distribution (SED) from radio to gamma-rays during the
rotation-powered state, shown in  the left panel of Fig.\ref{seds}.
We used the radio flux at 607\,MHz  as measured by \citet{Roy15} 
and the upper limits obtained at 1.4, 5.5 and 9\,GHz reported by \citet{Bassa14}. 
These are broadly consistent with the spectral slope expected from a radio pulsar 
with a power law index -1.7 \citep{Roy15}. 
For the UV/optical range we use the average U  and g', r', i' and J band
magnitudes obtained in 2014 and 2015 respectively, corrected for intervening
absorption $\rm E_{B-V}$ = 0.11 (dM10) and overlay the best fit blackbody
model with $\rm T_{bb} = 5200\pm$300\,K. The unabsorbed average 
X-ray spectra in the 0.3-10\,keV range 
of the two MOS cameras during Obs\,2  are reported together 
with the best fit power law with
$\Gamma$=1.07. We also add the {\it Fermi}-LAT
spectrum after the state transition from \citet{Xing14}, 
together with a power law with an exponential cut-off
with index $\Gamma$=1.8 and energy 
cutoff $\rm E_c = 2\,GeV$ obtained by the same authors. Here we note that
also  \citet{Johnson15} analyse the {\it Fermi}-LAT data obtaining comparable values. 
In the right panel of Fig.\ref{seds} we also report the SED obtained during
the previous LMXB state presented in dM13, substituting the 
{\it Fermi}-LAT fluxes, obtained at that time from the second release of the source
catalogue (2FGL),  with those reported by \citet{Xing14} during the same pre-transition
state together with their best fit power law exponential cut-off 
with index $\Gamma$=2.13 and 
$\rm E_c = 6\,GeV$. Also in this case the fit parameters 
are compatible within uncertainties
with those recently obtained by \citet{Johnson15} analysing the same period.


\noindent The SEDs at the two epochs substantially differ from each other
 implying very different contributions. 
In the post-transition state, the radio flux is a factor $\sim$6 lower than that 
observed in the disc state and the slope is consistent with that
of a radio pulsar \citep{Roy14}. The UV/optical/near-IR emission totally 
originates from the donor star, whilst an additional contribution
from the disc is required in the LMXB state (dM14). 
The X-ray emission in the radio pulsar state 
is harder than that before the transition ($\Gamma \sim$ 1.7, dM13). 
In the LMXB state the X-ray and gamma-ray emissions have 
been successfully modelled by a propeller scenario 
\citep{Papitto14,Papitto_Torres15} with marginal accretion from the NS, while
in the rotation-powered  state the X-ray emission 
mainly originates from the intrabinary shock.
The post-transition gamma-ray emission appears 
instead to originate in the pulsar outer magnetosphere
\citep{Johnson15} from the observed partially alignment of the {\it Fermi}-LAT
and radio spin light curves and the lack of an orbital modulation at GeV 
energies\footnote{The claim of a weak orbital modulation
by \citet{Xing14} during the rotation-powered state is not 
confirmed by \citet{Johnson15}}. 
The gamma-ray luminosity was found by \citet{Johnson15} to be  
$\sim 5\%$ of the pulsar spin-down energy.
Although the lack of X-ray coverage extending above 10\,keV cannot 
allow us to draw conclusions,
the present data suggest that the X-rays and high energy gamma-rays
are not linked.

\section{Conclusions}

We have analysed the X-ray and optical emissions of XSS\,J1227 during
the ongoing rotation-powered state into which the source entered at the
end of 2012/beginning 2013. We here summarize the main results:

\begin{itemize}
\item
The X-ray luminosity in the 0.3-10\,keV range 
during the radio pulsar state is $\rm \sim 1.0-1.7\times 10^{32}\,erg\,s^{-1}$, 
one order of magnitude lower than that in the
previous accretion-powered state. It is also found
to vary  by a factor $\sim$1.6 between two observations taken 6 months apart.
\item
We confirm the strong orbital X-ray and UV and optical modulations 
found by \citet{Bogdanov14}, \citet{Bassa14} and dM14, respectively.
\item
The X-ray modulation amplitude is found to vary by a factor of $\sim$2.7
betweem the two epochs. The variability is consistent with
X-rays originating at the intrabinary shock between the pulsar and donor star, 
which is eclipsed due to the companion orbital motion. Doppler boosting could be
important to shape the X-ray modulation when the source brightens. 
\item
The X-ray spectrum is non thermal ($\Gamma$=1.07) and it is harder than that
observed during the accretion-powered state and consistent with synchrotron
emission from the intrabinary shock. The spectrum also softens 
at orbital minimum, a new feature seen only in one other redback so far. 
This could be due to a softer
 contribution of the outer regions of the shock and/or of the NS. 
\item
The UV/optical/near-IR emission is modulated at the orbital period
due to irradiation of the companion star. 
We infer a 5500\,K  temperature for the unheated hemisphere of the donor and a
$\sim$6500\,K temperature for the irradiated face. We also derive 
a binary inclination  $46^o \lesssim i \lesssim 65^o$ and a mass ratio
$0.11\lesssim q \lesssim 0.26 $. For $\rm M_{NS} =1.4\,M_{\odot}$, 
the donor mass is $0.15 \lesssim \rm M_{2} \lesssim 0.36\,M_{\odot}$, 
confirming the redback  nature of this binary.
\item
The UV/optical orbital modulation also changes in amplitude with time. The 
simultaneous
UV and X-ray modulations behave opposite,  the UV flux and amplitude being 
larger when the X-ray flux  and amplitude are lower. This behaviour could be 
due to small changes in the mass flow rate
from the donor star which affect the size of the intrabinary shock.
\item
The broad-band SED from radio to gamma-rays in the disc-free state is different
from that observed during the LMXB state and composed by  multiple
contributions. The radio and GeV emissions are 
compatible with an origin from highly accelerated particles in the
pulsar magnetosphere, while 
the X-rays are dominated by the intrabinary shock. 
The UV/optical and near-IR fluxes instead  originate in the donor star. 
Whether the companion is heated by the intrabinary shock or by the pulsar
spin-down power needs to be assessed with a 
higher energy coverage in the X-rays.

\end{itemize} 

\section*{Acknowledgments}
This work is based on observations obtained with {\it XMM-Newton}, 
an ESA science mission with instruments and contributions directly 
funded by ESA Member States, under programme 0729560.
The authors wish to thank Dr. Norbert Schartel and the ESAC staff 
for their help in obtaining the XMM-Newton observation.
The REM observations were obtained under programme DDT-REM:30901 and 
AOT-31:31002. The REM team is acknowledged for the support 
in the scheduling and data delivery.
We acknowledge support from ASI/INAF I/037/12/0 and TMB from PRIN-INAF 2012-6.
The Barcelona group acknowledges support from the grants AYA\,2012-39303
and SGR\,2014-1073. We are grateful to the anonymous referee, whose comments 
improved the clarity of this work.
 
\bibliographystyle{mn2e}
\bibliography{xss_paperlow_accepted}

\vfill\eject

\end{document}